\documentstyle[12pt]{article}
\begin{document}
\begin{titlepage}
\setcounter{page}{0}

\vspace{2cm}
\begin{center}
{\Large HEAVY MAJORANA NEUTRINO DECAYS AND CP-PARITY VIOLATION}
\vspace{1cm}

{\large ALANAKYAN R.A., ASATRIAN H.M., IOANNISIAN A.N., CHATRCHYAN S.A.} \\
\vspace{1cm}
{\em Yerevan Physics Institute, Alikhanyan Br. 2, Yerevan, 375036, Armenia}\\
{\em e-mail: "asatryan@vxc.yerphi.am"}\\
{\em or: "alanak@vxc.yerphi.am"}\\
\end{center}

\vspace{5mm}
\centerline{{\bf{Abstract}}}
In  the  framework  of the left-right  symmetric   model  the
CP-parity violation has been studied in
heavy  Majorana neutrino
decays $N\rightarrow e^{\mp} \mu^{\pm} n$.
     The numerical estimates of CP-asymmetry for different  masses
of neutrino and $W_{R}^{\pm}$-boson were obtained. The possibility
to  detect this phenomenon at  high energy colliders is mentioned.

\vfill
\centerline{\large Yerevan Physics Institute}
\centerline{\large Yerevan 1995}

\end{titlepage}
\newpage

     Recently  the  heavy  Majorana  neutrino  production  and  their
further decays in $e^+e^-, ep, pp$ -collisions were actively studied
(see \cite{1,2} and references therein).
     Heavy Majorana neutrinos appear in many theories  beyond  the
Standard Model (massive Majorana  particles,  gluino,  neutralino,
gravitino, appear also in supergravity theories).
Their introduction helps to solve such important problems  as
the smallness of the left-handed neutrino mass (see the Appendix A
for the neutrino mass matrix),
baryonic  asymmetry  of  universe
and etc \cite{3,4,5,6}.
     On the other hand, the  introduction  of  Majorana  neutrinos
leads to some new phenomena such as the double  neutrinoless  $\beta$-decay
and other processes with lepton number violation.
Recently, in some works \cite{2,7} there have been shown that  CP-parity
violation can be induced in the decays of Higgs bosons
($H^o\rightarrow \overline{t} t, W^+ W^-, Z^o Z^o$)
by  heavy Majorana neutrino loops.

     The aim of  this  work  is  the  investigation  of  CP-parity
violation in heavy Majorana neutrino decays
$N\rightarrow e^{\mp} \mu^{\pm} n$.
It is well known that the existence of CP-violation requires the
presence of two phases.
In our case the CP violation arises due to the interference of two
diagramms (see Fig. 1). The resulting CP asymmetry is proportional
to the phases of Kobayashi-Maskawa like mixing matrix and the
imaginary part of one of the diagramms of Fig. 1, which is connected
with nonzero width of $W_{R}^{\pm}$- boson.
It should be mentioned  that this    phenomenon,
is  absent  in the
case of the Dirac neutrino decay, because in  this  case  the
decay is described by only one diagram.
    It should be noted that CP-parity violation was  investigated
also in $N\rightarrow e^+ e^- n $ decays \cite{8}.
It appeared due to the  interference of two
W-boson diagrams with the third diagram, also contributing to this
process and containing nondiagonal $Z^oNn$ vertex, absent  in  cases  with
Dirac neutrino. In case of the $N\rightarrow e^+ e^- n $ decay
the interference
of both W-bosonic  diagrams
with each other did not lead to CP-parity violation \cite{8}.

     We are working in  the  framework  of  left-right symmetric  model,
however our numerical results are  applicable  also  to  the  models
without $W_{R}^{\pm}$ e.g. to the Standard Model with
right neutrinos \cite{1}. In this case $N\rightarrow e^{\mp} \mu^{\pm} n$
decays  take place due  to  the exchange of standard $W_{L}^{\pm}$-bosons.

The total widths  of
processes $N\rightarrow e^{\pm} \mu^{\mp}n$ -
$\Gamma \equiv \Gamma(N\rightarrow e^- \mu^{+}n)$
and $\overline{\Gamma} \equiv \Gamma(N\rightarrow e^+ \mu^{-}n)$
described by Fig. 1 are equal to each other.
However, the differential widths, as it will be seen below, can be
different.
To compare the differential widths of these decays let us consider
the partialy integrated decay widths $\Gamma_{y_1> z}$,
$\bar\Gamma_{y_2> z}$, where
$y_{1,2}=\frac{2\epsilon_{1,2}}{m}$,
$\epsilon_{1,2}$
are energies of $e^-,\mu^+ (\mu^-,e^+)$ in
$N\rightarrow e^{-} \mu^{+}n,
(N\rightarrow e^+ \mu^{-}n)$
processes, m is mass of neutrino N. Thus $\Gamma_{y_1> z}$ is proportional
to the number of events $e^- \mu^{+}n$ in the decay
$N\rightarrow e^{-} \mu^{+}n$ with electron energy
$\epsilon_{e_{1}}>\epsilon_{o}$ and $\bar\Gamma_{y_2> z}$
is proportional to the numbers of events in the
decay $N\rightarrow e^+ \mu^{-}n$ with positron
energy  $\epsilon_{e_{2}}>\epsilon_{o}$, ($z=\frac{2\epsilon_{o}}{m}$).

Deriving Kobayashi-Maskawa (KM)  matrix  elements  in  the
$N\rightarrow e^{\mp} \mu^{\pm} n$
amplitude processes:
\begin{eqnarray}
&&M=V_{Ne}V_{n\mu}^\ast M_1-V_{N\mu}^\ast V_{ne}M_2 \nonumber\\  % (1)
&&\bar{M}=V_{N\mu}V_{ne}^\ast M_1-V_{Ne}^\ast V_{n\mu}M_2
\end{eqnarray}
after some simple calculations ($M_i, \overline{M_i}$ are
determined  in  formulae
(5), (6)) we have:
\begin{equation}
\Gamma_{y_1> z}-\bar\Gamma_{y_2> z}=2{\rm Im}\,(V_{Ne}V_{n\mu}  % (2)
^* V_{N\mu} V_{ne}^*)\frac{1}{4m}\int\limits_{y_1>z}\,
{\rm Im}\,(M_1M_2^+)\,d\Phi
\end{equation}
Here $d\Phi$ is the differential three-particle phase space.

For $z=0$ formula (2) results to the total widths difference, which
is equal to
zero in accordance with formula (B5) as we have mentioned above.
    From (2) one can see  that  the  effect exists only if
the factor ${\rm Im}\,(V_{Ne}V_{n\mu}^* V_{N\mu} V_{ne}^*)$
is different from zero. Indeed,  all
four elements of KM matrix in interaction  $Wen$   (formula  (A7)),
included  in  this  factor are  at  the  verteces  of
rectangles (formula (3) below), by the phase  transformations
of fermionic fields one can eliminate the phase only of  one line
and one column in V matrix (in formula (3)  they  are  marked  out  by
solid lines). Thus, at least one of above mentioned
KM matrix  elements
may have imaginary part.
\begin{equation}V=\left(
\begin{array}{l|l|lll}
\cdots&\quad \cdot&\cdots&\quad\cdot&\cdots\\
\hline                                                 %   (3)
\cdots&V_{Ne}&\cdots&V_{ne}&\cdots\\
\hline
\cdots&\quad\cdot&\cdots&\quad\cdot&\cdots\\
\cdots&V_{N\mu}&\cdots&V_{n\mu}&\cdots\\
\cdots&\quad \cdot&\cdots&\quad\cdot&\cdots
\end{array}
\right)
\end{equation}

     As it is seen  from  formula  (2),  for
the existing of CP-parity  violation
the phase difference between amplitudes $M_1$ and
$M_2$ (${\rm Im}M_1 M_2^+\neq 0$)
is  also  necessary.

   In \cite{9,10,11} CP-parity violation due the interference
of  the imaginary part of tree diagram with a real part of
loop diagram in t-quark decay $t\rightarrow d_i\bar{d_j}u$
was studied (see also \cite{12}, where CP violation appeared due to the
t-quark nonzero width).
In our case the effect appears due to the fact
that for one of diagrams $W_{R}^{\pm}$ boson is on mass
shell (as a result
the imaginary part of the W- boson  propagator occurs) and is virtual
on the second diagram and vise versa.

It must be noted that, the CP- violation
can arise also due to the interference of diagramms Fig. 2
(see also \cite{13}) with
the diagramms Fig. 1. Below  we will show (see formula (18) ) that
their contribution is relatively small.

Thus, we consider the range of masses:
\begin{equation}
m>m_{R}>m_{n}         % (4)
\end{equation}
where $m_{n}$ is the mass of the neutrino n.
     It should be noted \cite{9}, that we call diagrams 1 and 2  as  of
a tree type conditionally; each of them is the result of summation
of infinite number of diagrams  with  quark
and lepton loops. The finite width in the $W_{R}^{\pm}$-boson propagator
arises due to the imaginary parts of  these  loops
(see  formula (5), (6) below).

     The amplitudes $M_1, M_2$ may be written as follows:
\begin{eqnarray}
\nonumber M_{1}=\frac{g^2}{2}\bar{u}(k_1)\gamma_{\mu}P_{R}v(k)\bar{u}(k_{3})
\gamma_{\nu}
P_{R}v(k_{2})\frac{1}{m^2}\,\frac{1}{1-y_{1}-r_R^2+i\gamma} \times \\
\times \left[g_{\mu\nu}-
\frac{(k-k_{1})_{\mu}(k-k_{1})_{\nu}}{m_{R}^{2}}\right]
\end{eqnarray}                                                % (5)

\begin{eqnarray}
\nonumber M_{2}=\frac{g^2}{2}\bar{u}(k)\gamma_{\mu}P_{R}v(k_2)\bar{u}(k_1)
\gamma_\nu
P_{R}v(k_{3})\frac{1}{m^2}\,\frac{1}{1-y_2-r_R^2+i\gamma} \times \\
\times \left[g_{\mu\nu}-
\frac{(k-k_{2})_{\mu}(k-k_{2})_{\nu}}{m_{R}^{2}}\right]
\end{eqnarray}                                                 % (6)

    Here $r=\frac{m_{n}}{m}$, $r_R=\frac{m_R}{m},
\gamma=r_R\frac{\Gamma_R}{m}$,  symbols are introduced, $\Gamma_R$-
is the width of $W_R^{\pm}$-boson.
     Let us introduce the following  definition  of  CP-asymmetry:
\begin{equation}
A_{CP} = \frac{\Gamma_{y_1> z}-\bar\Gamma_{y_2> z}-\Gamma_
{y_1<z}+\bar\Gamma_{y_2< z}}{\Gamma_{y_1> z}+\bar\Gamma_
{y_2> z}+\Gamma_{y_1< z}+\bar\Gamma_{y_2< z}}=\frac{\Gamma_     % (7)
{y_1> z}-\bar\Gamma_{y_2> z}-\Gamma_{y_1> z}
+\bar\Gamma_{y_2> z}}{2\Gamma}.
\end{equation}

     While deriving formula (6) the obvious identities were  used:
\begin{eqnarray}
\nonumber \Gamma_{y_1> z}+\Gamma_{y_1< z}=\Gamma     \\
\bar\Gamma_{y_2> z}+\bar\Gamma_{y_2< z}=\Gamma.
\end{eqnarray}                                           %  (8)
Thus, we have the  difference  of
events $e^- \mu^{+}n$ with electron energies
$\epsilon_{e^{-}}>\epsilon_{o}$ and $e^+ \mu^{-}n$ with positron
energy  $\epsilon_{e^{+}}>\epsilon_{o}, z=\frac{2\epsilon_{o}}{m}$.

We can define more "symmetric" asymmetry
\begin{equation}
A_{CP}=\frac{\Gamma_{y_1> z}-\bar\Gamma_{y_2> z}-\Gamma_
{y_1< z}+\bar\Gamma_{y_2< z}-(\Gamma_{y_2> z}-\bar\Gamma_{y_1> z}-
\Gamma_{y_1<z}+\bar\Gamma_{y_2< z})}{4\Gamma}
\end{equation}                                            %  (9)
Using the formulae (1,2,5,6) one  can  show,  that  (7)  and  (9)
asymmetries are equal to each other.

     As a result of our calculations (Appendix B), we have:
\begin{equation}
A_{CP}=v f(r,r_R,z),
\end{equation}                                            %    (10)
where
\begin{equation}
v=\frac{{\rm  Im}(V_{Ne} V_{n\mu}^* V_{N\mu} V_{ne}^*)}
{|V_{Ne}|^2|V_{n\mu}|^2+ |V_{N\mu}|^2|V_{ne}|^2}
\end{equation}                                           %    (11)

     It is obvious, that $v \leq 0.5$.
     The  dependence of the function f on z
at various r, $r_{R}$ is shown on Figs 3-5. From Fig. 3-5 we see that
the effect is maximal for some middle values of z.

In our  calculations  we
assumed the main $W_R$-boson decay modes  to be   the  decays  in
hadrons $W_{R}\rightarrow ud$ as well as $W_{R}\rightarrow ln$ and so
\begin{equation}
\Gamma_R=3\Gamma_{ud}+\Gamma_{ln}
\end{equation}                                           %    (12)

     Let us estimate the possibility of the effect observation at
high energy colliders. Using the formula for
$W_{R}\rightarrow en$ decay widths \cite{14}
\begin{equation}
B(W_R\rightarrow en)\approx 0.04|V_{Ne}|^2\quad
\left[2+\frac{m_n^2}{m_R^2}\right](1-\frac{m_n^2}{m_R^2})
\end{equation}                                              %   (13)
and also (B10) we have:
\begin{equation}
B(N\rightarrow ne^-\mu^+)= 0.02\left[|V_{Ne}|^2|V_{N\mu}|^2+
|V_{Ne}|^2|V_{n\mu}|^2\right]\quad
\left[2+\frac{m_n^2}{m_R^2}\right](1-\frac{m_n^2}{m_R^2})
\end{equation}                                              %   (14)

     Hence, the difference of event number is:
\begin{eqnarray}
\Delta N_{CP}=
R (B(N \rightarrow e^{-} \mu^{+}n)_{\varepsilon_{e^{-}}>\varepsilon_{0}}
-B(N \rightarrow e^{+} \mu^{-}n)_{\varepsilon_{e^{+}}>\varepsilon_{0}} \\
\nonumber
-B(N \rightarrow e^{-} \mu^{+}n)_{\varepsilon_{e^{-}}<\varepsilon_{0}}  -
B(N \rightarrow e^{+} \mu^{-}n)_{\varepsilon_{e^{+}}<\varepsilon_{0}}) =
2R A_{{CP}}B(N \rightarrow e^{+} \mu^{-}n)
\end{eqnarray}                                                  %   (15)
Here R is the number of heavy neutrinos produced in colliders,
$R=2 \sigma \int {\cal L} dt $- for the
production of the pair of Majorana neutrino,
$R= \sigma \int {\cal L} dt $  -  for  production of the single
Majorana neutrinos.
Taking into account that $\sqrt{s} \gg m_{R},m_{N}$ \cite{1}
for the diagrams with t-chanel exchange of $W_{R}$ boson we have:
\begin{equation}
\sigma(e^+e^-\rightarrow NN)=|V_{Ne}|^4\times 100{\rm nbn}    % (16)
\left(\frac{m_L}{m_R}\right)^2
\end{equation}
{}From (16) one has for $m_R \sim 800GeV$ and integrated luminosites
${\cal L}
=3\cdot10^{41}s^{-1}$
number of neutrinos $R=6\cdot10^5|V_{Ne}|^4$.
Taking into account that $f \sim 0.1 \div
0.5$ for wide range of masses and middle values of z one has:
\begin{equation}
\Delta N_{CP}=(2.5\times10^3\div1.25\times10^4)r{\rm Im}
(V_{Ne}V_{n\mu}^* V_{N\mu} V_{ne}^*)|V_{Ne}|^{4}(1-\frac{m_n^2}{m_R^2})
\left\{2+\frac{m_n^2}{m_R^2}\right\}               % (17)
\end{equation}
     The heavy neutrinos may also be produced at the  peak  of
$Z^{\prime}$-
bosons.
     From \cite{15} (p.496) we have, that when
$m_{Z^{\prime}} = 750GeV$ and ${\cal L}=10^{34}cm^{-2}s^{-1}$,
the number of neutrino $R \sim 3\times 10^{7}$
(assuming $m_{Z^{\prime}} \gg 2m_{N}$ and so
$\Gamma(Z^{\prime} \rightarrow NN) \sim \Gamma(Z^{\prime} \rightarrow \mu^{+}
\mu^{-})$).
So, in decays of $Z^{\prime}$ the $\Delta N_{CP}$ is increased  at  least
by 2-3 orders.

It should be  also  noted,  that  our  results  can  be  also
applicable to the models with heavy Majorana  neutrino  without
$W^{\pm}_{R}$-bosons.
However, in these models $N\rightarrow e^{\mp} \mu^{\pm} n$ processes
occur due to  the  standard  $W_L^{\pm}$-
boson exchange, the  interaction of which   with  N  and  n  contains  the
additional smallness of order $\xi$ in comparison with
$W_{R}^{\pm}Ne$ interaction.

As mentioned above the considered effect of CP violation
arises due to the existence of nonzero width of
$W_{R}$- boson, which is the result of summation
of infinit number quark and leptons loop diagramms. That is why
we must estimate the contribution of other loop diagramms.
Let us consider the contribution of diagramms Fig. 2.
The contribution of diagramm Fig. 2a are GIM supressed by factor
$V_{Ne}V_{ne}^{*}\frac{m_{l}^{2}}{m_{N}^{2}}$ which arise from
internal charged lepton lines. More considerable
contribution comes from the interference of diagramm Fig. 2b
with diagramms of Fig. 1.
The CP -asymmetry due to this contribution is of order
\begin{equation}
A_{CP} \sim \frac{v \alpha^{2} r}{\sum B(W_{R}\rightarrow ln)} \sim
10^{-3}rv                 % (18)
\end{equation}
(this estimate is true when $m_{n}$ is not very close to $m_{R}$),
where $\alpha$ is fine structure constants.
As a result we see that for optimal z's
-----(Fig. 3-5) the contribution of the diagramm Fig. 2
is 100-10 times smaller of our result.

There exist the additional loop diagramms with Z-bosons and photon
exchange which can contribute to the considered process. Their
contribution is of the same order as diagramms of Fig. 2.

     We plan also to  study  CP  odd  correlations  in  decays  of
charged leptons, particularly, in
$\mu^{\pm}\rightarrow e^{\pm} \nu_{i}\nu_{j}$
decays, where  $\mu$-meson  and/or
electron are polarized. The work in this direction is presently in
progress.

     The authors are express their gratitude to  I.G.Aznauryan and
S. G. Grigoryan  for  fruitful
discussions.

     The research described in this publication was made possible in part
by Grant N MVU000 from the International Science Foundation.
\renewcommand{\theequation}{A.\arabic{equation}}
\setcounter{equation}0
\newpage
\appendix
\begin{center}
    APPENDIX A
\end{center}
\vspace{2cm}

     Spontaneous symmetry breaking of $SU(2)_L \times SU(2)_R \times U(1)$
leads  to  a
mass terms in the neutrino sector:
\begin{equation}
L_m = -\overline{\nu}_L m_D \nu_R - \overline{\nu}_R^cM \nu_R+h.c.
\end{equation}                                         %                 (A1)
     The Dirac ($m_D$) and Majorana (M) mass matrices are related  to  the
standard doublet and right Higgs triplet vacuum expectation:
\begin{eqnarray}
\nonumber m_D& =& f_{\nu}<\varphi>  \\
M& =& h<\chi>
\end{eqnarray}                                                    % (A2)
     One can obtain the connection  between  the  weak
eigenstates  $\nu_L$  and $\nu_R$ and the
Majorana mass eigenstates $\nu$, N as  a power  series
in $\xi = \frac{m_D}{M}$ (assuming $\det M \gg \det m_D$)  \cite{1}:
\begin{eqnarray}
\nu_L=P_L \nu+\xi P_LN+O(\xi^2) \\  %                          (A3)
\nu_R=P_R \nu+\xi^T P_R \nu+ O(\xi^2)                                  %
%%                  (A4)
\end{eqnarray}
where $P_{L,R}=(1\pm \gamma_5)/2$.
The masses of N and $\nu$ are
\begin{eqnarray}
M_N=M+O(\frac{1}{M})  \\
m_{\nu}=-m_D\frac{1}{M}m_D^T+ O(\frac{1}{M^3})
\end{eqnarray}                            %                         (A6)
Thus, requirement $<\varphi>\ll <\chi>$
naturally leads  to  large  Majorana mass
for  right-handed  neutrinos.  Left-handed  neutrinos  are  nearly
massless. Above requirement provides also large $W_R^{\pm}$-masses.
Interaction of the right-handed neutrinos with $W_R^{\pm}$-bosons
has the following form:
\begin{equation}
L= \frac{g}{\sqrt{2}}\overline{l}\hat{W}_RP_RVN +h.c.
\end{equation}             %  (A7)
In (7) V is a Kobayashi-Maskawa type mixing matrix in the leptonic part
of the charged current.
\renewcommand{\theequation}{B.\arabic{equation}}
\setcounter{equation}0
\newpage
\appendix
\begin{center}
    APPENDIX B
\end{center}
\vspace{2cm}

     For width difference of $N\rightarrow e^{\pm}\mu^{\mp}n$
processes we have:
\begin{eqnarray}
\nonumber
\Gamma_{y_1> z}-\overline{\Gamma}_{y_2> z}= {\rm Im}\,(V_{Ne} V_{N\mu}
^* V_{n\mu} V_{ne}^*)\frac{mg^{4}r}{512\pi^3}\times\\
\int\limits^{1+r^2}_{q(r,z)}dx \int\limits^{y_{+}}_{z}dy_{1}
{\rm Im}\left[\frac{1}{1-y_1-r_R^2+i\gamma}\times
\frac{1}{1-y_2-r_R^2-i\gamma}\right]
g(x,y_1,r,r_R),
\end{eqnarray}                                                          %  (B1)
\vspace{1cm}
where
\begin{equation}
g(x,y_1,r,r_R)=[(1+r^2-x)(4+r^2r_R^{-2})-
2r_R^{-2}y_1(1-y_1-r_R^2)-2r_R^{-2}y_2(1-y_2-r_R^2)] %(B2)
\end{equation}
\vspace{1cm}
\begin{equation}
y_{\pm}=\frac{1}{2}[2-x\pm
\sqrt{x^2-4r^2}]
\end{equation}
\vspace{1cm}
\begin{equation}
q(r,z)=\frac{(1-z)^2+r^2}{1-z}
\end{equation}
As a result of integration by $y_1$ one has:
\begin{eqnarray}
\nonumber
\Gamma_{y_1> z}-\overline{\Gamma}_{y_2> z}={\rm Im}\,(V_{Ne} V_{N\mu}
^* V_{n\mu} V_{ne}^*)\frac{rmg^4}{512\pi^3} \times \\
\int\limits_{q(r,z)}^{1+r^2}dx [a(r,r_R,z)c(r,r_R,z)]+ b(r,r_R,z)]
\end{eqnarray}
\vspace{1cm}
\begin{equation}
a(r,r_R,z)=  (1-x+r^2)(8+r^2r_R^{-2}-2r_R^{-2}x)-
4(1-r_R^2)(1-r^2r_R^{-2})-
4\gamma^2r_{R}^{-2}  \\
\end{equation}
\begin{eqnarray}
\nonumber c(r,r_{R},z) = (\arctan \frac{y_{+}+x-1-r_{R}^2}{\gamma}
-\arctan \frac{z+x-1-r_{R}^2}{\gamma} -\\
\arctan \frac{y_{+}-1+r_{R}^{2}}{\gamma}
+ \arctan \frac{z-1+r_{R}^{2}}{\gamma})
\frac{1}{x-2r_R^2}
\end{eqnarray}
\begin{eqnarray}
b(r,r_{R},z)=-2\gamma r_R^{-2}
(ln((y_{+}-1+r_{R}^2)^2+\gamma^2)-
ln((z-1+r_{R}^2)^2+\gamma^2)\\
\nonumber -ln((y_{+}+x-1+r_{R}^2)^2+\gamma^2)-
ln((z+x-1+r_{R}^2)^2+\gamma^2))
\end{eqnarray}
where $x=\frac{2\epsilon_{n}}{m}$, $\epsilon_{n}$
neutrino energy.

The witdh of the decay $N\rightarrow e^-\mu^+n$ is:
\begin{eqnarray}
\Gamma=\frac{mg^4}{512\pi^{3}}\left[|V_{Ne}|^2|V_{N\mu}|^2+
|V_{Ne}|^2|V_{n\mu}|^2\right]\int\limits^{1+r^2}_{2r}dx
\int\limits^{y_{+}}_{y_{-}}dy_{1}
\left[\frac{g(x,y_1,r,r_R)}{(1-y_1-r_R^2)^{2}+\gamma^{2}}
\right]    \\
\nonumber -\frac{rmg^4}{256\pi^{3}}{\rm Re}(V_{Ne} V_{N\mu}
^* V_{n\mu} V_{ne}^*)
\int\limits^{1+r^2}_{2r}dx
\int\limits^{y_{+}}_{y_{-}}dy_{1}
\left[\frac{g(x,y_1,r,r_R)}{x-2r_{R}^{2}}
\frac{1-y_{1}-r_{R}^{2}}{(1-y_1-r_R^2)^{2}+\gamma^{2}}
\right]
\end{eqnarray}
The first term of formula (B9) is the sum of squares of amplitudes
$M_1$, $M_2$
and in the  limit  of  $\gamma \rightarrow 0$  (i.e.  when
$m-m_R \gg \Gamma_R$,$m_R-m_n \gg \Gamma_R$)
will  give  the  main
contribution to the width:
\begin{equation}
\Gamma = \Gamma(N \rightarrow W_{R}^{+}e^{-})Br(W_{R}^{+}\rightarrow
n\mu^{+})+\Gamma(N \rightarrow W_{R}^{-}\mu^{+})Br(W_{R}^{-}\rightarrow
ne^{-})
\end{equation}
The second term in the formula (B9) is the  interference
of amplitudes and  for the
range of masses $m-m_R \sim \Gamma_R$, $m-m_R \sim \Gamma_R$
one can not ignore it in general.
     However, even in this range interference can be considerably
less than squares of amplitudes if the multiplier
$r{\rm Re}(V_{Ne} V_{N\mu}^* V_{n\mu} V_{ne}^*)$ in the interference
term in (B9) is  suffuciently less
than
$\left[|V_{Ne}|^2|V_{N\mu}|^2+|V_{Ne}|^2|V_{n\mu}|^2\right]$
which contains  the sum of the squared amplitudes of (B9).

\newpage
\begin{center}
Figure Captions
\end{center}
\vspace{2cm}
Fig.1 Diagramms contributing to the Majorana nentrino decays
$N\rightarrow e^-\mu^+n(N\rightarrow e^+\mu^-n)$.

Fig.2 Loop diagramms contributing to the Majorana nentrino decays
$N\rightarrow e^-\mu^+n(N\rightarrow e^+\mu^-n)$.

Fig.3 Function $f(r,r_R,z)$ versus $z$ at $r=0.8$.
Curves 1,2,3 denoted $r_R=o.95, 0,9, 0,85$ respectively.

Fig.4 Function $f(r,r_R,z)$ versus z at $r=0.6$. Curves 1,2,3,4 denoted
$r_R=0.7, 0.9, 0.8$ respectively.

Fig.5 Function $f(r,r_R,z)$ versus  z at $r=0.3$.
Curves 1,2,3,4 denoted $r_R=0.9,0.5,0.8,0.7$ respectively.
\end{document}